\documentclass[aps,twocolumn,prd]{revtex4-2}
\usepackage{amssymb,amsfonts,amsthm,amsmath,titlesec,graphicx,epsfig,subfig,color,bm,makeidx,fancyhdr,float,braket,simplewick,tikz,tikz-cd,appendix,mathrsfs,cancel,slashed,titletoc}

\usepackage{lipsum}
\usepackage[all]{xy}
\usepackage{dcolumn}
\usepackage[normalem]{ulem}
\usepackage[colorlinks,linkcolor=blue, urlcolor=blue, anchorcolor=blue, citecolor=blue]{hyperref}
\usepackage[font=bf]{caption}
\usepackage{newtxtext,newtxmath}

\def\nn{\nonumber}

\begin{document}
\title{Holographic dictionary from bulk reduction}
\author{Wen-Bin Liu}\email{liuwenbin0036@hust.edu.cn}
\author{Jiang Long}\email{longjiang@hust.edu.cn}
\affiliation{School of Physics,  Huazhong University of Science and Technology,\\
Luoyu Road 1037, Wuhan, Hubei 430074, China}

\begin{abstract}
We propose a holographic dictionary which comes from reducing the bulk theories in an asymptotically flat spacetime to its null infinity. A general boundary theory is characterized by a fundamental field, an infinite tower of descendant fields, constraints among the fundamental field and its descendants as well as a symplectic form. For the Carrollian diffeomorphisms, we can construct the corresponding Hamiltonians which are also the fluxes from the bulk, and whose quantum operators realize this algebra with a divergent central charge. This central charge reflects the propagating degrees of freedom and can be regularized. For the spinning theory, we need a helicity flux operator to close the algebra which relates to the duality transformation.
\end{abstract}
\maketitle

{\bf\emph{Introduction}}.---Holography, one of the main properties of gravity, has been checked for various examples in the context of AdS/CFT. However, most of the physical processes occur in an asymptotically flat spacetime. Flat holography \cite{Strominger:2013jfa,Strominger:2017zoo,He:2014cra,Pasterski:2017kqt,Donnay:2022aba,Bagchi:2022emh,Donnay:2022wvx,He:2014laa,Strominger:2014pwa,Hawking:2016msc,Pasterski:2016qvg}, which is to study the correspondence between the gravity in asymptotically flat spacetime and field theory at its boundary, has received a lot of attention in recent years. 


In this letter, we 
study the boundary theories from the bulk reduction and find the intrinsic properties of physically meaningful boundary theories. We provide a bulk-to-boundary dictionary which is summarized in table \ref{tab:my_label} and has been checked by virtue of unifying the results from the real scalar, Maxwell, and (linearized) gravity theories \cite{Liu:2022mne,Liu:2023qtr,Liu:2023gwa}. All the results will be extended to the higher spin theory \cite{Liu:2023jnc}, and we will show how the correspondences in this table are realized in a concise way which may also be true for general dimensions \cite{Li:2023xrr}.
\vspace{-2mm}
\begin{table}[h]
    \centering
    \begin{tabular}{|c|c|}  \hline\text{Bulk}&\text{Boundary}\\\hline\hline
   \text{AFS}&\text{Carrollian manifold}\\\hline
   \text{Leading radiative modes}&\text{Fundamental fields}\  $F_{A(s)}(u,\Omega)$\\\hline
   \text{Other modes}&\text{Descendant fields}\ \\\hline
   \text{EOMs}&\text{Constraints}\ $\mathcal{C}(F,F^{(k)})=0$\\\hline
   \text{Symplectic form}\ $\bm\Omega^{\mathcal{H}}(\delta {\tt f};\delta{\tt f})$&\text{Symplectic form} \ $\bm\Omega(\delta F;\delta F)$\\\hline
    \text{Leaky fluxes}&\text{Hamiltonians}\\\hline
   \text{$\#$ of Propagating DOFs}&\text{Proportion of central charges}\\ \hline
    \end{tabular}
    \caption{Bulk-to-boundary dictionary.}
    \label{tab:my_label}
\end{table}\vspace{-2mm}

{\bf Notations.} We introduce a four-dimensional null vector $n^\mu=(1,n^i)$ with $n^i$ the unit normal vector on the sphere, 
and then define $Y_{\mu\nu}^A=Y_\mu^An_\nu-Y_\nu^An_\mu$ with $Y^A_\mu=-\nabla^An_\mu=-\nabla^A\bar n_\mu$. 
The components of \(Y_{\mu\nu}^A\) give six conformal Killing vectors (CKVs) \(Y_{0i}^A,Y_{ij}^A\) on the sphere that satisfy 
$\nabla_AY_B+\nabla_BY_A=\gamma_{AB}\nabla_CY^C$. Similarly, we can define a complementary null vector $\bar n^\mu=(-1,n^i)$ and then $\bar{Y}_{\mu\nu}^A=Y_\mu^A\bar{n}_\nu-Y_\nu^A\bar{n}_\mu$.
We use the parameters \(f\), $g$ and $Y^A$ to label supertranslations, super-duality transformations, and superrotations, respectively. At last, the following abbreviation will be used a lot
\begin{align}
  \int dud\Omega=\int_{-\infty}^\infty du\int_{S^2}\sin\theta d\theta d\phi.
\end{align}

\vspace{5pt}
{\bf\emph{Boundary ingredients}}.---In this section, we discuss several ingredients at the boundary, including the manifold, symmetry, fields and constraints.

{\bf Manifold.} The bulk manifold is assumed to be asymptotically flat and a typical example is the Minkowski spacetime.  In retarded coordinates $x^\mu=(u,r,\theta,\phi)$, we may choose a timelike hypersurface $\mathcal{H}_r$ with constant $r$. By taking the limit $r\to\infty$ while keeping $u$ finite, the hypersurface $\mathcal{H}_r$ approaches future null infinity $\mathcal{I}^+$ with topology $\mathbb{R}\times S^2$. This is a Carrollian manifold \cite{Une,Gupta1966OnAA,Henneaux:1979vn} whose
  metric could be obtained by taking a Weyl scaling for the induced metric of the hypersurface $\mathcal{H}_r$ in the above limit. The metric of $\mathcal{I}^+$ is degenerate 
\begin{equation}
ds^2_{\mathcal{I}^+}=\bm\gamma=d\theta^2+\sin^2\theta d\phi^2
 \end{equation} which characterizes the unit sphere $S^2$. There should be a complementary null vector 
$\bm\chi=\partial_u$ that lies in the kernel of the metric $\bm\gamma$ and generates the retarded time direction.   \emph{Therefore, an asymptotically flat spacetime takes a Carrollian manifold as its boundary.}  

{\bf Symmetry.} Carrollian diffeomorphism \cite{Ciambelli:2018xat,Ciambelli:2019lap,Donnay:2019jiz}, which preserves the null structure of $\mathcal{I}^+$, is generated by the vector $\bm\xi$ such that the direction of the null vector $\bm\chi$ is invariant under Lie derivative along $\bm\xi$ \cite{Liu:2022mne}
 \begin{equation} 
\mathcal{L}_{\bm\xi}\bm\chi=\mu\bm\chi.\label{CD}
 \end{equation}
The general solution of \eqref{CD} is 
\begin{equation}
\bm\xi=\bm\xi_{f,Y}=f(u,\Omega)\partial_u+Y^A(\Omega)\partial_A\label{Cardif}
\end{equation} 
where $f$ is any smooth function on $\mathcal{I}^+$ and $Y^A$ is any smooth vector  on $S^2$. All the Carrollian diffeomorphisms consist of a group denoted by ${\rm Diff}(S^2)\ltimes C^\infty(\mathcal{I}^+)$, i.e., the semi-product of the diffeomorphisms on the sphere and the transformations generated by $f\partial_u$ with $f\in C^\infty(\mathcal{I}^+)$. 

{\bf BMS group.} The original Bondi-Metzner-Sachs (BMS) group \cite{Bondi:1962px,Sachs:1962wk,Sachs:1962zza} is a subgroup of the Carrollian diffeomorphism which is generated by $\bm\xi_{f,Y}$ with parameters satisfying
\begin{equation}
f(u,\Omega)=f(\Omega)+\frac{1}{2}u\nabla_A Y^A,\quad \Theta_{AB}(Y)=0
\end{equation} 
where the symmetric traceless tensor $\Theta_{AB}(Y)$ is defined as
\begin{equation}
\Theta_{AB}(Y)=\nabla_AY_B+\nabla_BY_A-\gamma_{AB}\nabla_CY^C.
\end{equation} 
The vanishing of the tensor $\Theta_{AB}(Y)$ indicates that $Y^A$ is a CKV. There are various extensions of the BMS group \cite{Barnich:2010eb,Barnich:2011mi,Campiglia:2014yka,Campiglia:2015yka,Campiglia:2015qka,Campiglia:2020qvc,Duval_2014a,Duval_2014b}. We will adopt the terminology defined in \cite{Liu:2023gwa} that a general supertranslation (GST) is generated by $\bm\xi_f=f(u,\Omega)\partial_u$, and a special superrotation (SSR) has generator $\bm\xi_Y=Y^A(\Omega)\partial_A$. The GST and SSR, consisting of the Carrollian diffeomorphisms, preserve the null structure of $\mathcal{I}^+$, so they are our main focus. 
We have also called the cases of $Y^A=Y^A(u,\Omega)$ a general superrotation  and of $f=f(\Omega)$ a special supertranslation.

{\bf Boundary fields.} For a bulk system whose fields are collected as ${\tt f}(t,\bm x)$, 
a variation of the Lagrangian ${\bf L}[{\tt f}]$ leads to 
\begin{equation}
\delta\mathbf L[{\tt f}]=\frac{\delta{\bf L}[{\tt f}]}{\delta {\tt f}}\delta {\tt f}-d\bm\Theta(\delta {\tt f};{\tt f})
\end{equation} where the first term represents the equation of motion (EOM) and the second term gives the presymplectic potential. To solve the EOM, we may expand the bulk field near $\mathcal{I}^+$
\begin{equation}
{\tt f}(t,\bm x)=\frac{F(u,\Omega)}{r}+\sum_{k=2}^\infty \frac{F^{(k)}(u,\Omega)}{r^k}
\end{equation} in the Cartesian coordinates. We have omitted the tensor indices and superscript  for the leading order field.
The coefficients $\{F(u,\Omega), F^{(k)}(u,\Omega)\}$ are boundary fields and the EOM becomes the following constraint equations
\begin{equation} 
\mathcal{C}(F,F^{(k)})=0
\end{equation} 
among the boundary fields. The boundary fields are classified into the fundamental fields and  descendant fields. The fundamental fields at $\mathcal{I}^+$ represent the leading radiative modes, namely 
\begin{align}
   {\tt f}_{A_1\cdots A_s}=r^{s-1}F_{A_1\cdots A_s}+\mathcal{O}(r^{s-2})
\end{align}
in retarded frame. The order of other components is at least $\mathcal{O}(r^{s-2})$.
We will write the fundamental fields as $F_{A(s)}$ for brevity \footnote{The fundamental fields for the theories with spin $s=0,1$, and $2$ can also be denoted by $\Sigma,\ A_A$ and $C_{AB}$, respectively, to match with the notations of the general literature.}, where $A(s)=A_1\cdots A_s$  represents a set of symmetric indices. 
Note that for the spin-2 theory, the field can also be interpreted as the metric perturbation $\delta g_{\mu\nu}=g_{\mu\nu}-\eta_{\mu\nu}$.

{\bf Constraints and descendants.}  The descendant fields, determined by the fundamental fields up to initial data, are not independent radiative  modes. For example, we consider the scalar theory with a potential $V(\Phi)=\sum_{n\ge4}\lambda_n\Phi^{n}/n$. From the bulk EOM $\partial^2\Phi-V'(\Phi)=0$, one can derive the constraints at order $\mathcal{O}(r^{-k-1})$ with $k\ge2$
\begin{align}
    \dot{\Sigma}^{(k)}=&-\frac{k-2}{2}\Sigma^{(k-1)}-\frac{1}{2(k-1)}\nabla_A\nabla^A\Sigma^{(k-1)}\\
    &+\frac{1}{2(k-1)}\sum_{n\ge4}^{k+2}\lambda_n\sum_{k_1,\cdots, k_{n-1}\ge1}^{k_1+\cdots +k_{n-1}=k+1}\Sigma^{(k_1)}\cdots \Sigma^{(k_{n-1})}.\nn
\end{align}
Descendant fields $\Sigma^{(k)}$ with $k\ge2$ are constrained and will be determined with appropriate initial conditions. A similar equation has been obtained in \cite{Satishchandran:2019pyc,Bekaert:2022ipg}.

For the Maxwell theory in the radial gauge $a_r=0$, the EOM at order $\mathcal{O}(r^{-k})$ with $k\ge 1$ can be worked out
\begin{align}
  n^\mu \dot{F}_{\mu\nu}^{(k)}+\frac{k-1}{2}(n^\mu+\bar{n}^\mu)F_{\mu\nu}^{(k-1)}+Y^{\mu A}\nabla_A F_{\mu\nu}^{(k-1)}=0, \label{consvector}
\end{align}
where the asymptotic field strength tensors in terms of boundary fields read
\begin{align}
    {F}_{\mu\nu}^{(k)}=&-Y_{\mu\nu}^A\dot{A}_A^{(k-1)}+(k-1)n_{[\nu}\bar n_{\mu]}A_u^{(k-1)}+Y_{\mu\nu}^A\nabla_A A_u^{(k-1)}\nn\\
    &-\frac{k-2}{2}(Y_{\mu\nu}^A+\bar Y^A_{\mu\nu})A_A^{(k-2)}+2Y_\mu^A Y_\nu^B\nabla_{[A}A{}^{(k-2)}_{B]}.\nn
\end{align}
Here $A_u^{(k)}$ with $k\ge1$ are asymptotic terms of $a_u$. In \eqref{consvector}, the equation with $k=1$ implies that the fundamental field $A_A$ is not constrained and the equations with $k\ge 2$ impose  constraints between $A_A$ and its descendants \footnote{Actually, we could also analysis the EOM for the Maxwell theory in the Lorenz gauge, which is natural to be generalized to the linearized gravity and higher spin theory. In \cite{Satishchandran:2019pyc}, the authors discussed the asymptotic expansion and EOM for the scalar, vector and gravitational theory in general dimensions with details. Interested readers can find a more systematic and robust method there. 
}. 

\emph{Therefore, the bulk equations of motion become constraints of boundary fields. The  fundamental fields represent leading radiative modes, while the descendant fields could be determined from fundamental fields given specified initial conditions.}

{\bf Symplectic form.} In the bulk, we may evaluate the symplectic form in a constant $r$ hypersurface $\mathcal{H}_r$
and find the boundary symplectic form \cite{Ashtekar:1981bq,Ashtekar:1981sf,Wald:1993nt,Iyer:1994ys,Wald:1999wa}  by sending it to $\mathcal{I}^+$ 
(with $32\pi G=1$ for the gravitational theory)
\begin{equation}
\bm\Omega_{s}(\delta F;\delta F)=\int du d\Omega \delta F^{A(s)}\wedge\delta\dot{F}_{A(s)},\label{symplform}
\end{equation} 
where the upper indices are raised by the inverse metric of $S^2$. 
From the boundary symplectic form, we could work out the commutators for the fundamental fields \footnote{We have also checked the fundamental commutators by virtue of the mode expansion of quantized fields, seeing \cite{Liu:2022mne,Liu:2023qtr,Liu:2023gwa} for details.} 
\begin{align}
    [F_{A(s)}(u,\Omega),\dot F_{B(s)}(u',\Omega')]=\frac{i}{2}  X_{A(s)B(s)}\delta(u-u')\delta(\Omega-\Omega'),
\end{align}
where the Dirac delta function on the sphere reads $\delta(\Omega-\Omega')=\frac{1}{\sin\theta}\delta(\theta-\theta')\delta(\phi-\phi')$. The tensor $X_{A(s)B(s)}$ is the symmetric and trace-free part of $\gamma_{A_1B_1}\cdots \gamma_{A_sB_s}$ with respect to two sets of indices $A(s)$ and $B(s)$, respectively. For the theories with spin $s=0$, we have $X=1$, while for $s=1$, we have $X_{AB}=\gamma_{AB}$. 


 One can also define vacua and then obtain the fundamental correlators
\begin{align}
    \bra0F_{A(s)}(u,\Omega)F_{B(s)}(u',\Omega')\ket0=i  X_{A(s)B(s)}\beta(u-u')\delta(\Omega-\Omega')\nn
\end{align}
with $\beta(u-u')=\int_0^\infty d\omega \frac{1}{4\pi\omega}e^{-i\omega(u-u'-i\epsilon)}$ \footnote{The function $\beta(u-u')$ is divergent, but its difference and time derivatives are finite. Moreover, it may be regularized by introducing infrared regulators, seeing \cite{Liu:2023qtr} for details.}, which will be useful for computing the central charges.

\vspace{5pt}
{\bf\emph{Hamiltonian operators and the algebras they generate}}.--- In this section, we will derive the Hamiltonians with respect to the GST and SSR at $\mathcal{I}^+$, and identify them as fluxes from the bulk. Then we will calculate commutators between the quantized Hamiltonian operators, which are shown to form a closed algebra realizing the aforementioned Carrollian diffeomorphisms.

{\bf Hamiltonians.} From the symplectic form \eqref{symplform} and the Carrollian diffeomorphism, we could make use of the formula below \cite{Iyer:1994ys,Wald:1999wa}
\begin{align}
    i_{\bm\xi}\Omega(\delta F,\delta F)=\delta H_{\bm\xi}
\end{align}
to find the corresponding Hamiltonian operators  \footnote{Actually, as shown in \cite{Liu:2023gwa}, up to some counterterms, the Hamiltonians are just the hard part of so-called BMS fluxes \cite{Compere:2018ylh,2020JHEP...10..116C,Compere:2020lrt,Donnay:2022hkf}  
which are derived in the full Einstein gravity. }
\begin{equation}
    \begin{aligned}
    &\mathcal{T}^s_f=\int dud\Omega\ f(u,\Omega) :\dot F_{A(s)}\dot F^{A(s)}:,\\
    &\mathcal{M}^s_Y=\frac{1}{2}\int dud\Omega\ Y_{A}(\Omega) P^{AB(s)CD(s)}\\
    &\hphantom{\mathcal{M}^s_Y=}\times (:\dot F_{B(s)} \nabla_CF_{D(s)}-F_{B(s)}\nabla_C \dot F_{D(s)}:),
\end{aligned}
\end{equation}
where we have imposed normal order to quantize the operators, and the tensor $P_{AB(s)CD(s)}$ is the symmetric and trace-free part of 
\begin{align}
    (\gamma_{AC}\gamma_{B_1D_1}+s\gamma_{AB_1}\gamma_{CD_1}-s\gamma_{AD_1}\gamma_{CB_1})\gamma_{B_2D_2}\cdots\gamma_{B_sD_s}
\end{align}
concerning $B(s)$ and $D(s)$, respectively. Note that the variation of field $\delta_{\bm\xi}F_{A(s)}$ needs to be modified to covariant variation, seeing \eqref{operatorform} and below.


{\bf Leaky fluxes.} Interestingly, the aforementioned Hamiltonians  are just the fluxes of Noether's charges arriving at \(\mathcal{I}^+\). Take the translation as an example. The four-momentum flux $\mathcal{T}^\alpha$ arriving at \(\mathcal{I}^+\), may be computed from the bulk stress tensor $T^{\mu\nu}$
\begin{align}
  {\cal T}^\alpha=\int_{\mathcal{I}^+} (d^3x)_rT^{\alpha r}=-\int dud\Omega\ n^\alpha \dot F_{A(s)}\dot F^{A(s)}.
\end{align}
One can check that the result is $\mathcal{T}_f$ with $f$ taking $-n^\alpha$.
Note that the quantity $T^\alpha$ is not a conservative charge but the leaky flux \cite{Wald:1999wa,Flanagan:2015pxa,2020JHEP...10..116C,Campiglia:2020qvc} from bulk to boundary. From the Poincar\'e fluxes, one can define two local density operators
\begin{subequations}
    \begin{align}
   &T^s(u,\Omega)=\ : \dot F_{A(s)}\dot F^{A(s)}:,\\
   &M^s_A(u,\Omega)=\frac{1}{2}(:\dot F^{B(s)} \nabla^CF^{D(s)}-F^{B(s)}\nabla^C \dot F^{D(s)}:)\nn\\
&\hphantom{M^s_A(u,\Omega)=}\times P_{AB(s)CD(s)}.
\end{align}
\end{subequations}
We can perform the (generalized) Fourier transforms to these density operators
\begin{subequations}
    \begin{align}
&\mathcal{T}^s_f=\int dud\Omega\ f(u,\Omega) T^{s}(u,\Omega),\\
    &\mathcal{M}^s_Y=\int dud\Omega\ Y^{A}(u,\Omega) M^s_A(u,\Omega).
    \end{align}
\end{subequations}
However, to preserve the null structure of $\mathcal{I}^+$ and get a closed algebra, we have to impose $\dot Y=0$ which makes  $\mathcal{M}^s_Y$  be the Hamiltonian operator related to the SSR. \emph{Hence, we conclude that the bulk leaky fluxes are precisely the corresponding Hamiltonians at the boundary.}

{\bf Actions on radiative fields.} We find that all the physical operators (including the  helicity flux operators below) have  the  unified form 
\begin{align}
  i\int dud\Omega :\dot F^{A(s)}\slashed\delta F_{A(s)}:\ .\label{operatorform}
\end{align}
where ``\(\slashed\delta F_{A(s)}\)'' denotes the corresponding covariant variation of $F_{A(s)}$ induced by taking the commutator with Hamiltonian operators. 
For example, the supertranslation operators can be written in the form below
\begin{align}
  \mathcal{T}_f^s=i\int dud\Omega :\dot F^{A(s)}\slashed\delta_fF_{A(s)}:\ ,
\end{align}
with the following covariant variation 
\begin{align}
  \slashed\delta_fF_{A(s)}\equiv i[\mathcal{T}_f^s,F_{A(s)}]=f\dot F_{A(s)}.
\end{align} 
It is easy to see that $\slashed\delta_fF_{A(s)}$ agrees with $\delta_fF_{A(s)}$ induced by Lie derivative \footnote{This $\bm\xi_f$ is the one extended to the bulk which has expansion with respect to radial coordinate $r$. Correspondingly, the Lie derivative acts on bulk field $f_{A(s)}$. In particular, for the gravitational theory, one needs to subtract the inhomogeneous terms in $\delta_fC_{AB}$ \cite{Liu:2023gwa},  which correspond to the soft part of fluxes at $\mathcal{I}^+$.}.

The covariant variation of superrotation is denoted as $\Delta_Y=\slashed\delta_Y-\slashed\delta_{f=\frac{1}{2}u\nabla\cdot Y}$, where the $\slashed\delta_Y$ corresponds to the standard superrotation \footnote{It is worth noting that the superrotations in our definition from Carrollian diffeomorphisms have removed the part like the supertranslations compared to the standard one.}.
Hence we  can define superrotation flux operators as
\begin{align}
  \mathcal{M}_Y^s=i\int dud\Omega :\dot F^{A(s)}\Delta_YF_{A(s)}:\ ,
\end{align}
where 
\begin{align}
&\Delta_YF_{D(s)}\equiv i[\mathcal{M}_Y^s,F_{D(s)}]\\
=&\rho_{AB(s)CD(s)}Y^{A}\nabla^{C}F^{B(s)} +\frac{1}{2}P_{AB(s)CD(s)}F^{B(s)}\nabla^{C}Y^{A},\nn
\end{align}
and we have defined $\rho_{AB(s)CD(s)}=\frac{1}{2}(P_{AB(s)CD(s)}+P_{AD(s)CB(s)})$.

Contrary to the supertranslation, the covariant variation of the superrotation differs from the one induced by Lie derivative. The origin lies in the fact $\delta_Y\gamma_{AB}=\Theta_{AB}(Y)\neq0$. To be adapted to the boundary metric, we introduce a ``connection'' $\Gamma_{AB}(Y)=\frac{1}{2}\Theta_{AB}(Y)$, and define the covariant variation $\slashed\delta_Y$ by regarding the variation induced by Lie derivative as the ordinary variation, in parallel to the definition of the covariant derivative. One can check the linearity, Leibniz rule, metric compatibility, and the last one, acting on scalar fields like the ordinary variation, i.e., \(\slashed\delta_Y\Sigma=\delta_Y\Sigma\).

\emph{In summary, we use the covariant variations to find a unified form for all the physical operators, which are adapted to the boundary metric, consistent with the starting point that all the fields locate at an asymptotically flat spacetime.}

{\bf Commutation relations.} We could write the general form of commutators among the supertranslation and superrotation operators for the theories with spin \(s\)
\begin{subequations}\label{algebra}
  \begin{align}
    &[\mathcal{T}^s_{f_1},\mathcal{T}^s_{f_2}]={\rm C}^s_T(f_1,f_2)+i\mathcal{T}^s_{f_1\dot f_2-f_2\dot f_1},\label{Vira}\\
    &[\mathcal{T}^s_f,\mathcal{M}^s_Y]=-i\mathcal{T}^s_{Y^A\nabla_A f},\\
    &[\mathcal{M}^s_Y,\mathcal{M}^s_Z]=i\mathcal{M}^s_{[Y,Z]}+is\mathcal{O}^s_{o(Y,Z)}.\label{mymz}
  \end{align}
\end{subequations} 
where we have defined the following function
\begin{align}
  o(Y,Z)=\frac{1}{4}\epsilon^{BC}\Theta_{AB}(Y)\Theta^A_C(Z),
\end{align}
 which vanishes by definition when $Y$ or $Z$ is a CKV. \({\rm C}^s_T(f_1,f_2)\) represent central charges which for \(s=0\) takes the form
\begin{align}
  {\rm C}^{(s=0)}_T(f_1,f_2)=-\frac{i\delta^{(2)}(0)}{48\pi}\int dud\Omega \left(f_1\dddot f_2-f_2\dddot f_1\right).\label{cenc}
\end{align}
The Maxwell field and gravitational field have two propagating degrees of freedom (DOFs) at $\mathcal{I}^+$, so their central charges are exactly twice larger than \eqref{cenc}. \emph{This fact illustrates that the number of propagating DOFs in the bulk corresponds to the proportion of the central charges at the boundary. This correspondence is true for general dimensions \footnote{In general $d$ dimension, the central charge for the Maxwell theory is $d-2$ times as large as \eqref{cenc}, as expected. For the spin-2 theory, the number of propagating DOFs is $d(d-3)/2$, which is exactly the proportion of this central charge compared to \eqref{cenc}. The reason lies in the symplectic structure whose form implies that $X_{A(s)B(s)}X^{A(s)B(s)}$ is exactly the number of propagating DOFs, and happens to be the proportion of ${\rm C}_T(f_1,f_2)$.}}.

 {\bf Regularization of $\delta^{(2)}(0)$.}  Using the orthogonal and complete relations for spherical harmonics $Y_{\ell,m}(\Omega)$, one can show that $\delta^{(2)}(0)$ is exactly the density of states on $S^2$, which is represented by an infinite sum  $\sum_{\ell,m}1$. One can use the spectral zeta function regularization \cite{1977CMaPh..55..133H,Elizalde:1994gf} for the Laplace operator on the compact manifold $S^2$ to find a finite result. One can also use the heat kernel method \cite{Polterovich1999HeatIO,Vassilevich:2003xt,1987PhRvD..36.3037B} to evaluate this sum. These two methods give the same results in $d=4$, namely $\delta^{(2)}(0)=1/12\pi$. These regularizations may be valid in higher dimensions \footnote{For more details about the regularization of the Dirac delta $\delta^{(d-2)}(0)$ on $S^{d-2}$, we refer readers to \cite{Li:2023xrr}.}.

{\bf Duality operators.} For the scalar theory, \eqref{algebra} realizes the Carrollian diffeomorphisms with a central extension. However, if the spin is not zero, we must include a generalized  duality operator \(\mathcal{O}_g^s\) to form a closed algebra, which can also be written in the aforementioned form \eqref{operatorform} by virtue of $\slashed\delta_gF_{A(s)}$.
As the result of $\dot Y=0$, we require the parameters $g$ of duality operators to be time-independent. The above covariant variations read 

\begin{align}
    [\mathcal{O}_g^s,F_{A(s)}]=-igQ_{A(s)B(s)}F^{B(s)},
\end{align}
 where $Q_{A(s)B(s)}$ is the symmetric and trace-free part of $\epsilon_{B_1A_1}\gamma_{A_2B_2}\cdots \gamma_{A_sB_s}$ with respect to $A(s)$ and $B(s)$, respectively.


As is known, the duality transformations is a symmetry for the free Maxwell theory \cite{Dirac:1931kp,Deser:1976iy,Bliokh:2012zr,Hamada:2017bgi,Hosseinzadeh:2018dkh,Seraj:2022qyt} and linearized gravity theory \cite{Henneaux:2004jw,Julia:2005ze,Bunster:2006rt}.  They rotate the field strength tensors with their Hodge duals. The infinitesimal duality transformations at $\mathcal{I}^+$ reduce to
\begin{align}
    \delta_\epsilon F_{A(s)}=\epsilon\widetilde F_{A(s)},\qquad  \delta_\epsilon \widetilde F_{A(s)}=-\epsilon F_{A(s)},
\end{align}
where $\widetilde F_{A(s)}$ is the dual fundamental field at $\mathcal{I}^+$, which relates to $F_{A(s)}$ through $\widetilde F_{A(s)}=Q_{A(s)B(s)}F^{B(s)}$. 
As an example, we could write out $\widetilde{A}_A=\epsilon_{BA}A^B, \ \widetilde C_{AB}=Q_{ABCD}C^{CD}$.

Now, one can derive the Hamiltonians related to the duality transformations and also the corresponding flux operators. The results are precisely $\mathcal{O}^s_g$, with parameters $g$ constants.  They can be extended to smooth functions on the sphere which we call special super-duality transformations (SSDTs), but can not be time-dependent due to the non-local term. Nevertheless, we call the latter as general super-duality transformation. The corresponding operators are called duality operators since it is the generator of duality transformations. Readers may need to be careful to distinguish this name from the dual charges constructed from the dual field $\widetilde F_{A(s)}$, such as dual mass/angular momentum, which are the dual counterparts to the normal charges \cite{Ramaswamy1981DualmassIG,Strominger:2015bla,Freidel:2018fsk,Godazgar:2018vmm,Godazgar:2018qpq,Godazgar:2018dvh,Seraj:2022qyt}. One can also refer to $\mathcal{O}^s_g$ as helicity flux operators since the fluxes evaluate the difference between the numbers of particles with left and right helicity. In particular, the flux is called optical helicity \cite{Bliokh:2012zr} for the Maxwell theory. 

It is amazing that the helicity flux operators appear in the commutators $[\mathcal{M}^s_Y,\mathcal{M}^s_Z]$. The duality transformation is supposed to have nothing to do with the rotation in the bulk. However, reducing to the boundary, the SSDT is intertwined with the SSR. The Hodge dual in the bulk reduces to exchanging the components of fields on $S^2$ at $\mathcal{I}^+$. 

{\bf Phase transformation.} Actually, when we switch the fundamental fields to complex scalars through the vector $\zeta^A=(1,-i/\sin\theta)$ and its complex conjugate $\bar\zeta^A$, namely taking $\psi^s=F_{A(s)}\zeta^{A_1}\cdots \zeta^{A_s}$ and $\bar \psi^s=F_{A(s)}\bar\zeta^{A_1}\cdots \bar\zeta^{A_s}$, the duality transformations become phase transformations of the complex scalars. The flux density operator corresponding to the phase transformation is \begin{align}
    O^s(u,\Omega)=-\frac{i}{2}(:\psi^s\dot{\bar{\psi}}^s-\bar\psi^s\dot\psi^s:),
\end{align} 
and we find the superrotation flux operator can be written as 
\begin{align}
    \mathcal{M}_Y^s=\int dud\Omega Y^AS^s_A+\frac{is}{4}\int dud\Omega o(y,\bar y)O^s(u,\Omega),
\end{align}
where 
\begin{align}
    S^s_A=\frac{1}{4}(:\dot{\bar\psi}^s\nabla_A\psi^s+\dot{\psi}^s\nabla_A\bar\psi^s-\bar\psi^s\nabla_A\dot{\psi}^s-\psi^s\nabla_A\dot{\bar\psi}^s:)
\end{align}
and 
\begin{align}
    o(y,\bar{y})=y\nabla_A\bar{\zeta}^A-\bar{\zeta}^A\nabla_A y-\bar{y}\nabla_A\zeta^A+\zeta^A\nabla_A\bar{y}
\end{align}
with $y=Y^A\zeta_A$ and $\bar y=Y^A\bar\zeta_A$. It implies that the superrotation flux operator can be divided into a duality part $\frac{is}{4}\mathcal{O}^s_{o(y,\bar y)}$ and a part which is in the same form as the superrotation flux operator in the spin-0 theory, namely that $\mathcal{S}^s_Y=\int dud\Omega Y^AS^s_A$ is just the form of the complex version of $\mathcal{M}_Y^{(s=0)}$. The duality part is proportional to the spin $s$ which leads to the appearance of $s$ in \eqref{mymz}. 

{\bf Intertwined algebra.} Including the duality operators for spinning theories, one needs to calculate out the commutators involving $\mathcal{O}^s_g$ to complete the algebra. The result is
\begin{subequations}\label{algo}
  \begin{align}
    &[\mathcal{T}^s_f,\mathcal{O}^s_g]=0,\\
    &[\mathcal{M}^s_Y,\mathcal{O}^s_g]=i\mathcal{O}^s_{Y^A\nabla_Ag},\\
    &[\mathcal{O}^s_{g_1},\mathcal{O}^s_{g_2}]=0.
  \end{align}
\end{subequations}
The algebra \eqref{algebra} and \eqref{algo} shows the intertwinement between Carrollian diffeomorphisms and SSDTs. 


{\bf Truncated algebras.}
Imposing $\dot Y=\dot g=0$ gives the previous closed algebra. 
If we further demand $Y^A$ to be a CKV, then we obtain the Newmann-Unti group \cite{Newman:1962cia,Barnich:2011ty} 
${\rm NU}(\mathcal{I}^+,\gamma,\chi)={\rm Conf}(S^2)\ltimes C^\infty(\mathcal{I}^+)$, where ${\rm Conf}(S^2)$ denotes the conformal transformations on the sphere.  
On the other hand, we could also require $\dot f=\frac{1}{2}\nabla\cdot Y,\dot Y=\dot g=0$, and the result is the generalized BMS group (intertwined with duality transformations for the spinning theory). 
At last, the original BMS group is obtained when $\dot f=\frac{1}{2}\nabla\cdot Y,\ Y^A=\omega^{\mu\nu} Y^A_{\mu\nu}$, and the Poincar\'e group  is 
recovered if $f=a^\mu n_\mu,\ Y^A=\omega^{\mu\nu} Y^A_{\mu\nu}$, where $a^\mu$ and $\omega^{\mu\nu}$ are constants.

{\bf The most general transformations}. We have calculated the most general algebra where $f$, $Y^A$, and $g$ are all smooth functions on $\mathcal{I}^+$. The resulting algebra takes the form
\begin{align}
        &[\mathcal{T}^s_{f_1},\mathcal{T}^s_{f_2}]={\rm C}^s_T(f_1,f_2)+i\mathcal{T}^s_{f_1\dot f_2-f_2\dot f_1},\nn\\
    &[\mathcal{T}^s_f,\mathcal{M}^s_Y]=i\mathcal{M}^s_{f\dot Y}-i\mathcal{T}^s_{Y(f)}+\frac{is}{2}\mathcal{O}^s_{\dot Y^A\nabla^B f \epsilon_{BA}}+\frac{i}{4}\mathcal{Q}^s_{\partial_u(\dot Y^A\nabla_A f)},\nn\\
    &[\mathcal{M}^s_Y,\mathcal{M}^s_Z]={\rm C}^s_M(Y,Z)+i\mathcal{M}^s_{[Y,Z]}+is\mathcal{O}^s_{o(Y,Z)}+{\rm N}^s_M(Y,Z),\nn\\
      &[\mathcal{T}^s_f,\mathcal{O}^s_g]=i\mathcal{O}^s_{f\dot g},\label{most}\\
  &[\mathcal{M}^s_Y,\mathcal{O}^s_g]={\rm C}^s_{MO}(Y,g)+i\mathcal{O}^s_{Y^A\nabla_A g}+{\rm N}^s_{MO}(Y,g),\nn\\
  &[\mathcal{O}^s_{g_1},\mathcal{O}^s_{g_2}]={\rm C}^s_O(g_1,g_2)+{\rm N}^s_O(g_1,g_2),\nn
\end{align}
which is valid for the higher spin bosonic field theory. 
We have introduced a new operator \(\mathcal{Q}^s_h=\int dud\Omega h(u,\Omega)\hspace{-1mm}:\hspace{-1mm}F^{A(s)}F_{A(s)}\hspace{-1mm}:\), 
which has no manifest physical meaning, since its action on fields is totally non-local. Therefore, we will not care about this operator, and it will disappear when \(\dot Y\) vanishes. Besides ${\rm C}^s_T(f_1,f_2)$, there are several central charges and non-local terms that will disappear for $\dot Y=\dot g=0$, and thus we will not give their explicit form here. Interested readers can find more details in \cite{Liu:2022mne,Liu:2023qtr,Liu:2023gwa}. These non-local terms come from the actions of  $\mathcal{M}^s_Y$ and $\mathcal{O}^s_g$ on fields, which obscure the interpretation of these operators. Recalling the above algebras truncated from \eqref{most}, one can use a schematic diagram to show how various groups are related:
\begin{widetext}
    \begin{align}
      \underset{\textstyle\text{most general case}}{f(u,\Omega),\ Y^A(u,\Omega),\ s} 
     \left\{\begin{array}{rll}
           \underset{\textstyle\text{no dual.}}{s=0} &
           \begin{cases}
               \underset{\textstyle\text{Carr. diffeo.}}{f(u,\Omega),\ \dot Y=0}\\[5mm]
              \underset{\textstyle\text{gener. BMS}}{\dot f=\tfrac{1}{2}\nabla\cdot Y,\ \dot Y=0}   
           \end{cases} & \\[1.2cm]
        \underset{\textstyle\text{with dual.}}{s\not=0} &
          \begin{cases}
              \underset{\textstyle\text{Carr. diffeo. $\times$ SSDTs}}{f(u,\Omega),\ \dot Y=\dot g=0}   \\[5mm]
              \underset{\textstyle\text{gener. BMS $\times$ SSDTs}}{\dot f=\tfrac{1}{2}\nabla\cdot Y,\ \dot Y=\dot g=0}  
          \end{cases} &
      \end{array}\right.\hspace{-9mm} \Rightarrow 
      \left\{\begin{array}{c}
             \underset{\textstyle\text{NU group}}{f(u,\Omega),\ Y^A=\omega^{\mu\nu}Y^A_{\mu\nu},\ s}\\[8mm]
             \underset{\textstyle\text{original BMS}}{\dot f=\tfrac{1}{2}\nabla\cdot Y,\ Y^A=\omega^{\mu\nu}Y^A_{\mu\nu},\ s}
          \end{array}\right. \hspace{-1mm}\Rightarrow
      \underset{\textstyle\text{Poincar\'e trans.}}{f=a^\mu n_\mu,\ Y^A=\omega^{\mu\nu} Y^A_{\mu\nu},\ s}\nonumber
    \end{align}
\end{widetext}

 Although we show a series of groups, it must be pointed out that our main concern is the Carrollian diffeomorphism. That is because it preserves the null structure of the $\mathcal{I}^+$ and leads to a closed algebra. The time-dependence of $f$ may be a little surprising. However, seeing $T^s(u,\Omega)=\ :\dot F_{A(s)}\dot F^{A(s)}:$ as a natural local object at $\mathcal{I}^+$ and taking the Fourier transform of this density implies that we need to consider the general $f$. Such a Fourier transform encodes all the information about the momentum radiation. In particular, if we take $f$ as a natural basis at $\mathbb R\times S^2$, i.e., $f=e^{-i\omega u}Y_{\ell,m}(\Omega)$, we find that \eqref{Vira} is a Virasoro algebra
for the real scalar, and the same is true for the spinning theories except that the central term will be twice as large. As a matter of fact, such a time-dependence appears in the context of the light-ray operator formalism  \cite{Kravchuk:2018htv,Cordova:2018ygx,Korchemsky:2021okt,Korchemsky:2021htm} which is related to the collider physics. 


{\bf Conclusions.} We propose a systematic way to obtain field theory at $\mathcal{I}^+$ from bulk reduction. We find some intrinsic characteristics of the boundary theory that capture the essence of the dictionary: \emph{There are fundamental fields that are unconstrained and determine their descendants up to initial data. The boundary symplectic forms only depend on these fields and give the fundamental commutators and correlators. The transformations of fundamental fields under the Carrollian diffeomorphism are adapted to the boundary metric and lead to the Hamiltonians or leaky fluxes at the boundary. Taking normal order results in quantum operators, which form a representation of the Carrollian diffeomorphism.} These properties have been checked by virtue of  the field theories with any integer spin,
which may provide new insight to the construction of the general Carrollian field theories. It will be our future interest to check the above properties of boundary theory, improve the dictionary, and investigate how these may help to understand the quantum gravity in the bulk. 

{\bf Discussions.} 
There are some open questions about the Carrollian diffeomorphism and flat holography.
\begin{enumerate}
    \item We consider the Carrollian diffeomorphism $\bm \xi_{f,Y}$ linking two boundary fields $F_1(u,\Omega)$ and $F_2(u,\Omega)$, whose bulk counterparts ${\tt f}_1(t,\bm x)$ and ${\tt f}_2(t,\bm x)$ are related by a bulk transformation. It is interesting to investigate how our boundary transformation will react the bulk physics or solution space in the future.
\item The BMS and its extensions are asymptotic symmetries which also reflect in the soft theorem of the bulk S-matrix. Allowing $f$ to depending on $u$, we are out of the soft limit. Accordingly, taking soft limit in the basis $f=e^{-i\omega u}Y_{\ell,m}(\Omega)$, the Carrollian diffeomorphism reduces back to the generalized BMS group for which the equivalence between the Ward identities and (leading and subleading) soft graviton theorems has been checked. It is natural to ask how the Carrollian diffeomorphism is integrated into S-matrix.  Under a Carrollian diffeomorphism generated by $\bm\xi$, there is an equation relating the fluxes $Q_{\bm\xi}$ at $\mathcal{I}^+$ and $\mathcal{I}^-$
\begin{align}
    Q_{\bm\xi}\Big|_{\mathcal{I}^+}-Q_{\bm\xi}\Big|_{\mathcal{I}^-}=\frac{1}{2}\int_{\text{bulk}}d^4 x T^{\mu\nu}\delta_{\bm\xi}g_{\mu\nu}\label{37}
\end{align}
like (4.77) in \cite{Li:2023xrr}. Bracketing \eqref{37} by in and out states, we will get a similar expression as Ward identity in \cite{Strominger:2013jfa} which relates to the soft theorem, except that there will be non-trivial contributions from the insertion of stress tensor in S-matrix. This problem deserves further study. 
\end{enumerate}

\vspace{3pt}
{\bf Acknowledgments.} 
The work of J.L. is supported by NSFC Grant No. 12005069.

\bibliography{refs}
\bibliographystyle{utphys}
\end{document}